# Effect of transition layers on the electromagnetic properties of composites containing conducting fibres

D. P. Makhnovskiy, L. V. Panina, D. J. Mapps

*Department of Communication, Electronic and Electrical Engineering, Plymouth University, Drake Circus, Plymouth, Devon PL4 8AA, United Kingdom.*

A. K. Sarychev

*Physics Department, New Mexico State University, Box 30001, Las Cruces, NM 88003-8001, USA.*

**Abstract** For the first time the approach to calculating the effective dielectric and magnetic response in bounded composite materials is developed. The method is essentially based on the renormalisation of the dielectric matrix parameters to account for the surface polarisation and the displacement currents at the interfaces. This makes it possible the use of the effective medium theory developed for unbounded materials, where the spatially-dependent local dielectric constant and magnetic permeability are introduced. A detailed mathematical analysis is given for a dielectric layer having conducting fibres with in-plane positions. The surface effects are most essential at microwave frequencies in correspondence to the resonance excitation of fibres. In thin layers (having a thickness of the transition layer), the effective dielectric constant has a dispersion region at much higher frequencies compared to those for unbounded materials, exhibiting a strong dependence on the layer thickness. For the geometry considered, the effective magnetic permeability differs slightly from unity and corresponds to the renormalised matrix parameter. The magnetic effect is due entirely to the existence of the surface displacement currents.

**Key words:** composite materials, effective parameters, effective-medium theory, mean-field theory, boundary effects, surface effect, transition layers, spatial dispersion.

## 1. Introduction

Metal-dielectric composite materials have received much attention because of their importance in modern technology.[1] Metallic inclusions, in particular metallic fibres, can reinforce the dielectric and magnetic properties of ceramics and plastic materials. Electromagnetic properties of the composite materials are analysed customarily in terms of the effective macroscopic parameters: dielectric constant $\varepsilon_{ef}$ and magnetic permeability $\mu_{ef}$, which are calculated by averaging the responses from material constituents.[2-6] The effective medium theory offers quick insight into linear problems, which are difficult to analyse by other means. However, it has disadvantages typical to all mean-field theories since it ignores the fluctuations in a system. It assumes that the local electric and magnetic fields are the same in the volume occupied by each component of a composite. In some cases the local field fluctuations by no means can be ignored, as in the case of a percolation composite in the frequency range corresponding to the plasmon resonances in metal grains.[7] Then, the application of EMT is rather questionable and adequate modifications are needed. Another example is bounded composites or composites containing interfaces. The microscopic local fields near the surfaces (or interfaces) differ considerably from those in the internal regions, due to the existence of the scattering fields from boundaries. In the approach developed here a specific surface polarisation is introduced into the EMT approach. The effect of the surface polarisation can be strong in thin materials, the characteristic size of which is smaller than the correlation length.[8] For elongated inclusions in the form of a fibre, their length $l >> r_0$, where $r_0$ is the fibre radius, corresponds to the correlation length. For many engineering materials $l$ is of the order of few mm, whereas the composite layer thickness is in the range of fractions of mm. In this case, the boundary effects must be taken into account when considering the effective response from a thin system. This is the purpose of the present paper.

A general approach to systems containing interfaces is solving the Maxwell equations in the regions, which are regarded as homogeneous and imposing the boundary conditions at the interfaces. The system is then characterised by $\varepsilon_{ef}$ and $\mu_{ef}$ having a stepwise variation. However, since the microscopic fields near the boundaries are different, the effective parameters vary gradually within certain transition regions (known as transition layers) adjacent to the interfaces. They also depend on properties and geometry of the media near both sides of the interface. In certain cases, the transition layers can change the response from the entire system even if the system is thick.[9,10] For example, the concept of a transition layer is used to explain the elliptical polarisation of the light reflected from an isotropic medium.[11,12] In the present work we also discus the effect of the transition layers on the dielectric response (calculating the reflection and absorption coefficients) from thick composite systems.

The composite materials with elongated conducting inclusions have a number of characteristics specific for this system, which eventually result in strong boundary effects. In such composites, the concentration of the percolation threshold is proportional to the aspect ratio $p_c \propto r_0 / l$.[6,13,14] In the limit $r_0 / l << 1$, $p_c$ is very small, however, the inclusion contribution to the effective dielectric constant becomes large already for very small concentrations $p << p_c$. It implies that the interaction between the fibres is strong even for $p << p_c$ and the assembly of conducting fibres is a system with a long-range strong interaction having a characteristic dimensionless correlation length of $l / 2r_0$. Then, the surface effects have to be essential at large distances comparable with the correlation length. On the other hand, the fibre currents induce surface displacement currents resulting in the fibre depolarisation, which weakens the interaction in the transition layer, changing the basic property of the system.

Another characteristic feature of composites containing conducting fibres is the existence of a manifold resonance at microwave frequencies ($\lambda_m \sim 2l$, where $\lambda_m$ is the wavelength in the dielectric matrix). This effect is responsible for the dispersion of the effective dielectric constant, which otherwise at these frequencies appears only in the vicinity of $p_c$.[6] For $p << p_c$, the frequency behaviour of $\varepsilon_{ef}$ is of the form [15]

$$\varepsilon_{ef}(\omega) = \varepsilon + \sum_n \frac{A_n}{\omega_{res,n}^2 - \omega^2 + i\beta_n \omega} \tag{1}$$

where $\varepsilon$ is the dielectric constant of the matrix and $\omega_{res,n}$ are the resonance frequencies. The interaction with the boundary changes the resonance excitation condition and the system has a different sequence of the resonance frequencies.[8] Therefore, the result of interaction with boundaries is the modification of dispersion for $\varepsilon_{ef}$ at microwave frequencies. In this analysis, the desperation characteristics are restricted to this frequency range (but $l_m >> r_0$).

The approach developed in this paper essentially uses a single-particle model within which the boundary effects are considered. First, the current distribution is analysed in the antenna approximation ($\ln l / r_0 >> 1$, $l_m >> r_0$) at the fibre placed in a dielectric layer of thickness *h*. This problem was investigated in Ref. 8, however, we need a more detailed analytical analysis which is of great importance for the calculation of $\varepsilon_{ef}$. The equation for current distribution in bounded layers can be transformed to the form valid for an infinite system using a renormalisation procedure. In this sense, the boundaries can be eliminated, instead of them new renormalised matrix parameters (dielectric constant $\varepsilon_b$ and magnetic permeability $\mu_b$) appear in the equation determining current at the fibre. This approach allows the further use of EMT in its form developed for unbounded materials. There are a number of approximations in the literature for obtaining the effective parameters of the composite materials with elongated inclusions. We use here the theory developed in Ref. 6 as the most complete and consistent with experiment and not restricted to the quasistatic limit. An important feature of this theory is a spatial dependence of the effective dielectric constant near a fibre for scales smaller then the fibre length. The nonlocal property of $\varepsilon_{ef}$ is even enhanced by the interaction with the boundary: the environment near the fibre is characterised by $\varepsilon_b$ and $\mu_b$ which depend on the layer thickness and the fibre position. As a result, the effective dielectric constant exhibits a strong dependence on the layer thickness for thin materials (thinner than the transition layer). In the case of thick materials, however, the role of the transition layers is not essential and boundaries effects can be neglected.

For the geometry considered the magnetic permeability $\mu_b$ arising in the renormalisation method constitutes the effective magnetic properties. It is originated by the combination of the fibre current and surface displacement current which together form a circulatory current inducing a magnetic moment directed along the magnetic field (real part of $\mu_b$ is larger than unity). In the present case, $(m_b - 1) << 1$ and is not noticeable in the experiments. However, if the composite layer is placed on the metal substrate it may be essential to take account of the magnetic response. The experimental observation of the magnetic response would be a direct confirmation of the boundary effects.

## II. Single-particle model

This section concerns a single-particle approximation [8] for the response of thin dielectric layer with a conductive fibre excited by an external electric field. This problem allows the effective parameters of layered composite materials to be found taking the boundary effects. The construction of a correct EMT equation uses the differential equation for current density at the fibre, which involves the effect of boundaries.

A model depicted in Fig. 1 is considered. A conducting fibre with a radius $r_0$ and a length $l$ is placed in a dielectric layer of thickness $h$ parallel to its surface at a depth $h_0$ ($r_0 < h_0 < (h - r_0)$). The layer is characterised by a dielectric constant $\varepsilon$ and magnetic permeability $\mu$. The fibre is assumed to be an ideal conductor. This approximation is reasonable when considering the boundary effects since the current distribution inside a fine fibre does not alter the polarisation at the surface. The system is subjected to an ac uniform electrical field directed along the fibre (*x*-direction): $\mathbf{E}_0 = (\mathrm{E}_{0x}, 0, 0)\exp(i\omega t)$. The *z*-direction is chosen along the normal to the layer, and the fibre is at $z = -h_0$, $y = 0$. The problem is considered in so called antenna approximation [16] ($l >> r_0$, $l_m >> r_0$ where $l_m = 2p\,c/w\sqrt{e\,m}$ ) which allows the distribution of the current density at the fibre to be represented as $\mathrm{j}(x)\delta(y)\delta(z+h_0)$ where $\delta$ is the Dirac function. The electric and magnetic fields $\mathbf{E}$ and $\mathbf{H}$ scattered by the fibre are convenient to write in terms of a vector potential $\mathbf{A}$ and scalar potential $\varphi$ (Gaussean units are used)

$$\mathbf{E} = -\frac{4\pi}{c^2} i\omega\mu\mathbf{A} + \frac{4\pi}{i\omega\varepsilon}\,\mathrm{grad}\,\mathrm{div}\mathbf{A}\,,\ \mathbf{H} = \frac{4\pi}{c}\mathrm{rot}\mathbf{A}\,. \qquad (2)$$

In the present case, only two components of the vector-potential $\mathrm{A}_x$ and $\mathrm{A}_z$ are needed, which are represented in the form of convolutions with current density $\mathrm{j}(x)$ (see Appendix A)

$$\mathrm{A}_{x1,2,3}(x,y,z) = \left(\mathrm{G}_{x1,2,3}(x,y,z) * \mathrm{j}(x)\right),\ \mathrm{A}_{z1,2,3}(x,y,z) = \left(\mathrm{U}_{1,2,3}(x,y,z) * \mathrm{j}(x)\right) \qquad (3)$$

The convolution of two functions $\mathrm{g}(x)$ and $\mathrm{f}(x)$ is determined as the following integral

$$(\mathrm{g}(x) * \mathrm{f}(x)) = \int_{-l/2}^{l/2} \mathrm{g}(x-s)\,\mathrm{f}(s)\,ds\,,$$

where indexes 1, 2, 3 designate areas $z \in [0, \infty +]$, $z \in [-h, 0]$ and $z \in [-\infty, -h]$, respectively. Function $\mathrm{G}_{x1,2,3}$ satisfies the equation

$$\Delta \mathrm{G}_{x1,2,3} + k_{1,2,3}^2\, \mathrm{G}_{x1,2,3} = \vartheta_{1,2,3} \qquad (4)$$

with the boundary conditions

$$\mu \mathrm{G}_{x2}\big|_{z=0} = \mathrm{G}_{x1}\big|_{z=0},$$
$$\mu \mathrm{G}_{x2}\big|_{z=-h} = \mathrm{G}_{x3}\big|_{z=-h},$$
$$\left.\frac{\partial \mathrm{G}_{x2}}{\partial z}\right|_{z=0} = \left.\frac{\partial \mathrm{G}_{x1}}{\partial z}\right|_{z=0},$$
$$\left.\frac{\partial \mathrm{G}_{x2}}{\partial z}\right|_{z=-h} = \left.\frac{\partial \mathrm{G}_{x3}}{\partial z}\right|_{z=-h}.$$

Here $k_{1,3} = \omega/c$ and $k_2 = (\omega/c)\sqrt{\varepsilon\mu}$ are the wave numbers in free space and dielectric layer, respectively, $\vartheta_{1,3} \equiv 0$, $\vartheta_2 = -\delta(x)\delta(y)\delta(z+h_0)$, and $\Delta = \partial^2/\partial x^2 + \partial^2/\partial y^2 + \partial^2/\partial z^2$ is the Laplace operator.

Functions $U_{1,2,3}$ satisfy the equation

$$\Delta U_{1,2,3} + k_{1,2,3}^2 U_{1,2,3} = 0 \tag{5}$$

with the boundary conditions

$$\varepsilon U_2\big|_{z=0} = U_1\big|_{z=0},$$
$$\varepsilon U_2\big|_{z=-h} = U_3\big|_{z=-h},$$
$$\left(\frac{\partial U_1}{\partial z} - \frac{\partial U_2}{\partial z}\right)\Bigg|_{z=0} = \left(\frac{1}{\varepsilon\mu} - 1\right)\frac{\partial G_{x1}}{\partial x}\Bigg|_{z=0},$$
$$\left(\frac{\partial U_3}{\partial z} - \frac{\partial U_2}{\partial z}\right)\Bigg|_{z=-h} = \left(\frac{1}{\varepsilon\mu} - 1\right)\frac{\partial G_{x3}}{\partial x}\Bigg|_{z=-h}.$$

Equation (4) together with the boundary conditions is self-sufficient, whereas the boundary conditions for equation (5) contain functions $G_{x1}$ and $G_{x3}$ entering the boundary conditions for equation (4).

For further analysis, only $G_{x2}$ and $\partial U_2/\partial z$ are needed

$$G_{x2}(x,y,z) = \frac{\exp(-ik_2 r)}{4\pi r} + \frac{1}{4\pi}\int_0^{+\infty} \frac{a_{x2}(k,h,h_0)\exp(\gamma_2 z) + b_{x2}(k,h,h_0)\exp(-\gamma_2 z)}{\Delta_2(k,h)\gamma_2} J_0(k\rho)k\,dk, \tag{6}$$

$$\frac{\partial U_2(x,y,z)}{\partial z}\Bigg|_{\varepsilon\neq\mu} = \frac{\partial}{\partial x}\left[\frac{1}{2\pi}\int_0^{+\infty} \frac{\left(a_{z2}(k,h,h_0)\exp(\gamma_2 z) - b_{z2}(k,h,h_0)\exp(-\gamma_2 z)\right)(\varepsilon\mu - 1)}{\Delta_1(k,h)\Delta_2(k,h)} \gamma_2 J_0(k\rho)k\,dk\right] \equiv \frac{\partial \tilde{U}_2}{\partial x} \tag{7}$$

where $r = \sqrt{x^2 + y^2 + (z+h_0)^2}$, $\rho = \sqrt{x^2 + y^2}$, $\gamma_1(k) = \sqrt{k^2 - k_1^2}$, $g_2(k) = \sqrt{k^2 - k_2^2}$ and $J_0$ is the Bessel function. In (7) a new function $\tilde{U}_2$ is introduced. Integrals in (6) and (7) use the following functions

$$\Delta_1(k,h) = \left(\gamma_2^2 + \gamma_1^2\varepsilon^2\right)\mathrm{sh}(\gamma_2 h) + 2\gamma_1\gamma_2\varepsilon\,\mathrm{ch}(\gamma_2 h),$$

$$\Delta_2(k,h) = \left(\gamma_2^2 + \gamma_1^2\mu^2\right)\mathrm{sh}(\gamma_2 h) + 2\gamma_1\gamma_2\mu\,\mathrm{ch}(\gamma_2 h),$$

$$\mathrm{a}_{x2} = (\gamma_2 - \gamma_1\mu)[\gamma_1\mu\,\mathrm{sh}(\gamma_2(h-h_0)) + \gamma_2\,\mathrm{ch}(\gamma_2(h-h_0))],$$

$$\mathrm{b}_{x2} = \exp(-\gamma_2 h)(\gamma_2 - \gamma_1\mu)[\gamma_1\mu\,\mathrm{sh}(\gamma_2 h_0) + \gamma_2\,\mathrm{ch}(\gamma_2 h_0)],$$

$$\begin{aligned}\mathrm{a}_{z2} = {} & [\gamma_1\mu\,\mathrm{sh}(\gamma_2(h-h_0)) + \gamma_2\,\mathrm{ch}(\gamma_2(h-h_0))](\gamma_2 + \gamma_1\varepsilon)\exp(\gamma_2 h) - \\ & - [\gamma_1\mu\,\mathrm{sh}(\gamma_2 h_0) + \gamma_2\,\mathrm{ch}(\gamma_2 h_0)](\gamma_2 - \gamma_1\varepsilon)\end{aligned},$$

$$\begin{aligned}\mathrm{b}_{z2} = {} & [\gamma_1\mu\,\mathrm{sh}(\gamma_2(h-h_0)) + \gamma_2\,\mathrm{ch}(\gamma_2(h-h_0))](\gamma_2 - \gamma_1\varepsilon)\exp(-\gamma_2 h) - \\ & - [\gamma_1\mu\,\mathrm{sh}(\gamma_2 h_0) + \gamma_2\,\mathrm{ch}(\gamma_2 h_0)](\gamma_2 + \gamma_1\varepsilon)\end{aligned}.$$

A special method of calculating the integrals in equations (6) and (7) is given in Appendix B.

The tangential component $\mathrm{E}_\tau$ of the total electric field at the fibre surface equals zero. In the approximation used the circulatory currents in the fibre are neglected, then this condition is written in the form

$$\mathrm{E}_\tau \approx (\mathrm{E}_{0x} + \mathrm{E}_x)\Big|_{\substack{x\in[-l/2,l/2] \\ y^2+(z+h_0)^2=r_0^2}} = 0 \qquad (8)$$

where $\mathrm{E}_x$ is the longitudinal component of the scattered field. Equations (2) and (8) give

$$-\mathrm{E}_{0x} = \mathrm{E}_x = \frac{4\pi}{i\omega\varepsilon}\left[k_2^2\,\mathrm{A}_{x2} + \frac{\partial^2 \mathrm{A}_{x2}}{\partial x^2} + \frac{\partial^2 \mathrm{A}_{z2}}{\partial x \partial z}\right]_{\substack{x\in[-l/2,l/2] \\ y^2+(z+h_0)^2=r_0^2}} \qquad (9)$$

Using equations (7), (9) and equality $\partial^2 A_{z2}/\partial x\,\partial z = \partial^2(\mathrm{U}_2 * \mathrm{j})/\partial x \partial z = \partial^2(\tilde{\mathrm{U}}_2 * \mathrm{j})/\partial x^2$ the basic integro-differential equation is obtained

$$\left[\frac{\partial^2}{\partial x^2}\left((\mathrm{G}_{x2} + \tilde{\mathrm{U}}_2) * \mathrm{j}\right) + k_2^2(\mathrm{G}_{x2} * \mathrm{j})\right]_{\substack{x\in[-l/2,l/2] \\ y^2+(z+h_0)^2=r_0^2}} = -\mathrm{E}_{0x}\frac{i\,\omega\,\varepsilon}{4\pi} \qquad (10)$$

For equation (10), the boundary conditions for current density at the fibre ends have to be imposed $\mathrm{j}(-l/2) = \mathrm{j}(l/2) = 0$.

Equation (10) is simplified calculating the convolutions approximately. Using equations (6) and (7), the following approximations for the functions $G_{x2}$ and $\tilde{U}_2$ are obtained in two limiting cases ($h \to 0$ and $h \to \infty$)

$$\lim_{h\to 0} \tilde{U}_2 = \frac{(\varepsilon\mu - 1)}{\mu} \frac{e^{-ik_1 r}}{4\pi r}, \qquad \lim_{h\to\infty} \tilde{U}_2 = 0, \tag{11}$$

$$\lim_{h\to 0} G_{x2} = \frac{1}{\mu} \frac{e^{-ik_1 r}}{4\pi r}, \qquad \lim_{h\to\infty} G_{x2} = \frac{e^{-ik_2 r}}{4\pi r}$$

Because the fibre has very small diameter $2r_0$, from equation (11) it follows that the real parts $\mathrm{Re}(G_{x2})$ and $\mathrm{Re}(G_{x2} + \tilde{U}_2)$ have sharp positive picks of the order of $1/r_0$ in the vicinity of $x = 0$ for any value of $h$. To the contrary, the imaginary parts $\mathrm{Im}(G_{x2})$ and $\mathrm{Im}(G_{x2} + \tilde{U}_2)$ are limited when both $x$ and $r_0$ approach zero. Integration of the real parts in the vicinity of the picks gives the main contribution to the integrals. In this case the convolutions are approximated as [17]

$$\left(\mathrm{Re}(G_{x2} + \tilde{U}_2) * j\right) \approx j(x) \int_{-l/2}^{+l/2} \mathrm{Re}(G_{x2} + \tilde{U}_2)\, dx = j(x) Q_1, \tag{12}$$

$$\left(\mathrm{Re}(G_{x2}) * j\right) \approx j(x) \int_{-l/2}^{+l/2} \mathrm{Re}(G_{x2})\, dx = j(x) Q_2$$

where the "form factors" $Q_1$ and $Q_2$ are positive and represent the area under the corresponding narrow bell-shaped curves.

Substituting approximations (12) into (10) yields

$$Q_1 \frac{\partial^2}{\partial x^2}\left( j(x) + i \frac{\left(\mathrm{Im}(G_{x2} + \tilde{U}_2) * j\right)}{Q_1} \right) + k_2^2 Q_2 \left( j(x) + i \frac{\left(\mathrm{Im}(G_{x2}) * j\right)}{Q_2} \right) \approx -\frac{i\omega\varepsilon}{4\pi} E_{0x} \tag{13}$$

Since the parameters $Q_1$ and $Q_2$ involve a large factor $2\ln(l/r_0)$ (see Appendix C), equation (13) can be further simplified when the convolutions with the imaginary parts are neglected in comparison with those with the real parts (see Appendix C)

$$Q_1 \frac{\partial^2 \varphi(x)}{\partial x^2} + k_2^2 Q_2 \, \varphi(x) \approx -\frac{i\omega\varepsilon}{4\pi} E_{0x} \qquad (14)$$

$$\varphi(-l/2) = \varphi(l/2) = 0$$

Equation (14) is a basic differential equation in terms of which the boundary effects are introduced in EMT with the aim to modify the effective parameters.

Equation (14) describes the current distribution in the layered system and allows a generalised expression for the resonance wavelengths $\lambda_{res}$ to be obtained. In an infinite medium, the value of $\lambda_{res}$ is given by [18,19]

$$\lambda_{res,n} = \frac{2l\sqrt{\varepsilon\mu}}{2n-1}; \; n = 1, 2, 3...$$

The effect of the boundaries results in a non-linear dispersion equation

$$\lambda_{res,n} = \frac{2l\sqrt{\varepsilon\mu}}{2n-1} \sqrt{\frac{Q_2(\lambda_{res,n}, h, h_0)}{Q_1(\lambda_{res,n}, h, h_0)}}; \; n = 1, 2, 3... \qquad (15)$$

The dependencies on thickness $h$ of the main resonance frequency $f_{res} = c/\lambda_{res,1}$ $(n = 1)$ obtained from (15) for different $\varepsilon$ ($\mathrm{Im}\,\varepsilon = \mathrm{Im}\,\mu \equiv 0$) are given in Fig. 2(a). The parameters used for all the calculations are $h_0 / h = 0.5$, $r_0 = 0.004$ mm, $l = 8$ mm. The resonance frequency changes from the vacuum value of $c/2l$ (18.75 GHz for $l = 8$ mm) at $h \to 0$ to the value of $c/2l\sqrt{\varepsilon}$ at $h \to \infty$ corresponding to that for an infinite medium with the dielectric constant $\varepsilon$. The characteristic feature of all the curves is the existence of two regions defined by the parameter $h_c$. For $h < h_c$, the resonance frequency rapidly drops with increasing $h$, and for $h > h_c$ it decreases slowly reaching the saturation limit. The meaning of the introduced parameter $h_c$ can be understood from Fig. 2(b), where the derivatives $\partial f_{res} / \partial h$ as functions of $h$ are given. For small $h$, this parameter has a constant large gradient, decreasing linearly with $h$. The value of $h_c$ is defined by continuing this line until it intersects zero derivative level (infinite medium). Then $h_c$ is a characteristic thickness when the system becomes sensitive to its outer boundaries as far as an electromagnetic response is concerned. For the case of Fig. 2, $h_c \approx 0.2$ mm. It is shown that $h_c$ is independent of the material parameter $\varepsilon$, but it depends on geometry of the inclusions being a function of the fibre length $l$ and radius $r_0$. The numerical analysis shows that this dependence is of a logarithmic type: $h_c \propto \ln(l/r_0)$, which can be associated with the energy stored in the fibre $L_0\, j^2/2$ where $L_0 \cong 2\ln(l/r_0)$ is the inductance per a unit length of a thin wire. The effect of the fibre position inside the layer is not essential for thin layers ($h < h_c$). The change in parameters of thick materials with respect to the fibre position is about 25%, as will be shown below.

To demonstrate the consistency of the model, the results obtained for the resonance frequency are compared with the available experimental data.[8] The experimental method [20-23] is based on measurements of the reflection coefficient from a composite system placed near a metal substrate. As the distance between the sample and substrate is increased the reflection signal exhibits an interference minimum at a frequency corresponding to the resonance frequency. The composite system consisted of a dielectric matrix containing aluminium-coated glass fibres with the volume concentration of about 0.02%. The metal coating is not always continuous and the effective concentration can be even smaller. This concentration is considerably smaller than the percolation threshold $p_c \sim 0.1\%$, allowing the single-particle approximation to be used for analysis. The dielectric matrix comprises a polymer with a metal powder, which makes it possible to reach large $\varepsilon$ with a small absorption ($\mathrm{Im}\,\varepsilon << \mathrm{Re}\,\varepsilon$). Figure 3 compares the theoretical plots with the experimental data (taken from Ref. 8) for the resonance frequency as a function of thickness *h*. For $\varepsilon = 52$ (obtained for the dielectric matrix from independent measurement), the experimental data can be fitted well by calculations for small *h*, and there is a large difference between the two curves for $h > h_c$, since the experiment does not come down to the limit corresponding to infinite medium with $\varepsilon = 52$. It can be related to the fine dispersed structure of the composite matrix used in Ref. 8 (polymer and metal powder) and the difficulty of determining the effective dielectric constant $\varepsilon$ near the fibre in this case. Another reason of this discrepancy can be related to a layered structure of thick samples as they are obtained by combining a number of 0.1-0.2 mm layers. The gaps between very thin layers can apparently decrease the effective value of $\varepsilon$ resulting in an increasing resonant frequency $f_{res} = c/2l\sqrt{e}$ . Further experimental analysis is needed to clarify this case. However, it is more important for our purpose that there is a good agreement for thin layers $h \le h_c$ which proves that the model describes functionally well the resonance properties of composites containing elongated inclusions in thin layers.

A considerable boundary effect results in the existence of surface layers where the effective dielectric constant is different from that in the inner region. Figure 4 shows the resonance frequency as a function of inclusion position $h_0$ for a thin layer ($h = 0.2$ mm) and thick layer ($h = 2$ mm). There is a transition layer $h_s$ ($h_s < h_c$ for $h < h_c$ and $h_s = h_c$ for $h > h_c$) within which the value of $f_{res}$ decreases and approaches that for the case when the inclusion is placed sufficiently inside the sample. It is seen that the variation in $f_{res}$ due to change in $h_0$ is considerably smaller than that when $h$ is altered. Thin layers ($h < h_c$) can be considered as resonantly uniform, in which $f_{res}$ is nearly a function of $h$ only and the dependence on $h_0$ can be neglected. In the case of $h > h_c$, this dependence is essential within the surface transition layers and the sample, generally speaking, can not be treated as uniform. However, as it will be demonstrated later, the existence of the surface transition layers does not alter the resonance frequency in thick samples. This means that they are not important in determining the effective dielectric constant. On the contrary, the effective medium theory (EMT) in thin composite materials with elongated inclusions has to be essentially modified due to strong boundary effects.

## III. EMT for thin composite layers below the percolation threshold ($h < h_c$, $p < p_c$)

There are a number of methods developed in the literature to calculate the effective macroscopic parameters of composite materials with non-spherical, in particular, elongated inclusions. However, they are entirely restricted to the case of unbounded materials. Our objective is to obtain an extension to the case of thin composite materials where the boundary effects can not be ignored. If the layer thickness is comparable with the size of inclusions embedded in it the concept of dielectric constant seems to lose its direct meaning. Then, there is a question if such a layer can be characterised by the effective parameter $\varepsilon_{ef}$. In the case under consideration, for $h < h_c$, all the fibres are subjected to near-same boundary influence, as was demonstrated in Fig 4. In this context, such a sample can be treated as a uniform layer characterised by effective parameters, yet, these parameters will depend on thickness $h$ and, in general, on properties of the surrounding media.

For our analysis, a modified EMT equation developed in Ref. 6 is used. This approach has been distinguished from other theories since it has a number of advantages: it gives a correct value of the percolation threshold and can be expanded to a non-quasistatic case when the boundary effects are most essential. Along with this, it enables technically to take into account the boundary polarisation. For this purpose, it is important that the method combines the Bruggeman EMT [2,24,25] and Maxwell-Garnet theory [26,27] with its idea that the local medium near different inclusions may be different. Within this approach it is possible to introduce a renormalised parameters $\varepsilon_b$ and $\mu_b$ for the medium near fibres as the result of interaction with the boundaries. These parameters will then be involved in EMT equations for $\varepsilon_{ef}$ and $\mu_{ef}$ and the boundaries can be omitted.

The single particle approximation considered above can be used for calculating the renormalised parameters. For this purpose, Eq. (14) describing the current distribution at the inclusion is transformed to a certain canonical form. In the antenna approximation ($l >> r_0$, $l_m >> r_0$) the current density $\mathrm{j}(x)$ at the fibre in the unbounded medium with some material constants $\varepsilon$ and $\mu$ is determined by Eq. (10) with $\tilde{\mathrm{U}}_2 \equiv 0$

$$\left[\frac{\partial^2}{\partial x^2}(\mathrm{G} * \mathrm{j}) + k^2 (\mathrm{G} * \mathrm{j})\right]_{\substack{x \in [-l/2, l/2] \\ y^2 + (z+h_0)^2 = r_0^2}} = -\mathrm{E}_{0x} \frac{i \omega \varepsilon}{4\pi}$$

This equation can be further simplified in a similar way as that used for obtaining equation (14), which requires $\ln(l/r_0) >> 1$

$$Q\frac{\partial^2 \varphi(x)}{\partial x^2} + k^2 Q\,\varphi(x) \approx -\frac{i\omega\varepsilon}{4\pi}E_{0x} \qquad (16)$$

Here $k = (\omega/c)\sqrt{\varepsilon\mu}$ is the wave propagation parameter, the function $\mathrm{G}(x,y,z) = \exp(-ikr)/4\pi r$ is related to the solution of Eq. (4) when $h = \infty$, and $Q = \int_{-l/2}^{+l/2} \mathrm{Re}(\mathrm{G})dx$. The form of Eq. (16) corresponding to the unbounded medium will be called a canonical form. In the case of bounded materials, Eq. (14) can be reduced to this canonical form by renormalising material parameters as

$$\varepsilon_b = \varepsilon Q/Q_1,\ \mu_b = \mu Q_2/Q,\ k_b = (\omega/c)\sqrt{\varepsilon_b\mu_b} = (\omega/c)\sqrt{\varepsilon\mu(Q_2/Q_1)}. \qquad (17)$$

This procedure allows the boundaries to be eliminated when determining the current distribution at the inclusion. They are replaced by a 'new' medium with $\varepsilon_b$, $\mu_b$, which appears near the inclusion at a characteristic distance $h_c$, as shown in Fig. 5. For $h < h_c$ the factors $Q_1$ and $Q_2$ depend weakly on $h_0$, then $\varepsilon_b$ and $\mu_b$ may be considered to be functions of thickness $h$ only. The appearance of the permeability $\mu_b > 1$ owes its origin to the magnetic moment related to the current at the fibre and the induced displacement current at the layer surface.

Figure 6 shows the plots of the renormalised parameters $\varepsilon_b$ and $\mu_b$ as functions of thickness $h$ at the resonance wavelength with $\varepsilon = 52$, $\mu = 1$. The value of $\varepsilon_b$ equals 1 for small $h \to 0$ and increases to its bulk value $\varepsilon$ when $h$ tends to infinite. The magnetic permeability $\mu_b$ differs slightly from 1 going through a maximum. This is related to the induced magnetic moment as a function of $h$: it tends to be zero in two limiting cases $h = 0$ and $h = \infty$. In the experiment when the metal substrate is placed away from the sample the magnetic response is not noticeable. In the case of the composite placed on the metal substrate the magnetic properties may become noticeable to be measured.

After we have calculated the renormalised parameters characterising the medium near fibres, the EMT equation can be constructed similar to the case of unbounded media. This equation uses the condition that the total polarisation averaged over all the inclusions has to be zero.[6,23] For bounded materials, the surface polarisation due to surface displacement currents has to be included as well, which gives

$$p P_{fibre} + (1-p) P_{matrix} + P_{surface} = 0 \quad (18)$$

where $p$ is the fibre concentration. The terms in Eq. (18) correspond to averaged polarisation of fibres, dielectric matrix and surface, respectively. Using renormalisation procedure (17), the surface contribution can be taken into account by means of the parameters $\varepsilon_b$ and $\mu_b$ which are used as renormalised matrix constants to determine the fibre polarisation $P_{fibre}$. In this approach $P_{surface} = 0$, however, the regions with parameters $\varepsilon_b$ and $\mu_b$ appear near fibres; their polarisation is different from that of the matrix and has to be included in Eq. (18).

The polarisation $P_{fibre}$ can be calculated from the current distribution on a fibre given by equation (13). It involves the dielectric constant of the surrounding medium $\varepsilon$. In the effective medium approach, $\varepsilon$ has to be replaced by $\varepsilon_{ef}$. However, the effective medium near elongated inclusions can not be considered uniform on a characteristic scale of the order of inclusion size *l*, and the corresponding effective parameter, $\tilde{\varepsilon}_{ef}$ depends on scale $s$. Since the total electric field on the inclusion is near zero, the interaction between the inclusions has little effect on the dielectric properties of the medium in the vicinity of them. Therefore, the value of $\tilde{\varepsilon}_{ef}(s)$ in this region equals $\varepsilon_b$ which differs from $\varepsilon$ as the result of boundary effects. Far from the inclusions at distances larger than *l*, the effective medium can be considered as uniform having the dielectric constant $\tilde{\varepsilon}_{ef} = \varepsilon_{ef}$ of a bulk material. It means that equation (13) obtained for a fibre in a uniform medium has to be modified. At this stage, it is also important to consider a fibre with a finite conductivity, which influences dispersion of the effective parameter via the skin effect. The equation for the current distribution accounting for the finite conductivity and scale-dependent $\tilde{\varepsilon}_{ef}(s)$ was developed in Ref. 6 for the case of infinite medium. It turns out that the calculation of fibre polarisation involves scaling along the fibre only. Assuming a linear scaling results in

$$\tilde{\varepsilon}_{ef}(\mathbf{r}-\mathbf{r}') = \varepsilon + 2(\varepsilon_{ef} - \varepsilon)\,|\mathbf{r}-\mathbf{r}'|/l \qquad (19)$$

where the points $\mathbf{r}$ and $\mathbf{r}'$ are taken on the fibre surface. In the case of the bounded composite the matrix constant $\varepsilon$ must be replaced by the normalised effective constant $\varepsilon_b$. Considering, that fibres are placed in the *x*-direction, (19) becomes

$$\tilde{\varepsilon}_{ef}(x-x') = \varepsilon_b + 2(\varepsilon_{ef} - \varepsilon_b)(x-x')/l \qquad (20)$$

Using the scale-dependence (20) the fibre polarisation can be calculated[6]

$$P_{fibre} = \frac{4 p i s^*_f / e_{ef} w}{1 + (4 p i s^*_f / e_b w)(4r_0^2/l^2)\ln(1 + l e_b / r_0 e_{ef})\cos\Omega_b} E_0 \qquad (21)$$

$$\Omega_b^2 = l^2 k^2 L_b C_b / 4,\ L_b = 2\ln(l/r_0) + i k_b l,\ C_b = e_b / \ln(1 + l e_b / r_0 e_{ef})$$

where $\Omega_b$ is the normalised resonance frequency, $L_b$ is the fibre inductance per unit length of fibre, and $C_b$ is the capacitance per unit length calculated after taking the account of the scale-dependence (21). The normalised fibre conductivity $\sigma^*_f = f(\Delta)\sigma_f$ takes into account the skin effect in the conducting fibre where $\sigma_f$ is the fibre conductivity. The function *f* is obtained from a classical skin-effect in a conducting cylinder [28]

$$f(\Delta) = ((1-i)/\Delta)\, J_1((1+i)\Delta) / J_0((1+i)\Delta)$$

where $J_0$ and $J_1$ are the Bessel functions, $\Delta = r_0\sqrt{2\pi\sigma_f \omega}/c$ is the ratio of the fibre radius $r_0$ to the skin depth $\delta = c/\sqrt{2\pi\sigma_f \omega}$.

The matrix is represented as an assembly of fine spherical particles of dielectric constant $\varepsilon$ which are embedded in the effective medium with dielectric constant $\varepsilon_{ef}$. The polarisation of the dielectric matrix is given by the standard quasistatic equation [28]

$$P_{matrix} = \frac{3(\varepsilon - \varepsilon_{ef})}{2\varepsilon_{ef} + \varepsilon} E_0 \qquad (22)$$

where $\varepsilon$ is the initial constant of the matrix (not $\varepsilon_b$ ). In general, the polarisation of the regions with $\varepsilon_b$ near fibres has to be considered separately from $P_{matrix}$. It seems reasonable to assume that they can be represented by ellipsoids with short axes $h_c/2$. The concentration of such 'new' inclusions is enlarged by a factor of $(h_c / 2r_0)^2$. With increasing $p$ these areas may change the matrix properties entirely. Here we consider that $p$ is sufficiently small to omit their contribution.

Substituting (21) and (22) into (18) yields

$$\frac{p}{2} \frac{4 p i s^*_f / e_{ef} w}{1 + (4 p i s^*_f / e_b w)(4r_0^2 / l^2) \ln(1 + l e_b / r_0 e_{ef}) \cos\Omega_b} + (1-p)\frac{3(e - e_{ef})}{2e_{ef} + e} = 0 . \qquad (23)$$

Here the factor 1/2 in the first term results from averaging by directions in the plane.

Equation (23) describes the effective response from a thin composite sample ( $h < h_c$ ) below the percolation threshold ( $p < p_c$ ).

**IV. Analysis of the effective response near resonance frequency**

In this part the dispersion of the effective response near the main resonance frequency is analysed. The constant $\varepsilon_{ef}$ is calculated from non-linear equation (23) for two concentrations $p = 2 \cdot 10^{-4}$ and $p = 2 \cdot 10^{-5}$. These small values of $p$ correspond to the composite materials used for the experimental investigation of $\varepsilon_{ef}$ in Ref. 8. Figure 7 shows the dispersion of the real ( $\varepsilon_1$ ) and imaginary ( $\varepsilon_2$ ) parts of $\varepsilon_{ef} = \varepsilon_1 + i\varepsilon_2$ for a very thin sample with $h = 0.05\,\text{mm} << h_c$. The frequency behaviour is of a resonance type with the resonance frequency corresponding to a maximum of imaginary part. In the case $p << p_c$, this frequency depends weakly on the inclusion concentration (see Fig. 7(b)). Figure 8 compares the dispersion characteristics of $\varepsilon_{ef}$ for layers of different thickness $h$. The case of an infinite system is also given. The main resonance frequency $f_{res} = c/2l\sqrt{\varepsilon_b}$ $(\mu_b \approx 1)$ is considerably shifted to the high frequency region since the renormalised effective constant $\varepsilon_b$ is several times smaller than the matrix constant $\varepsilon = 52$ for $h < h_c$. Besides, the next resonance which is clearly seen in the case of $\varepsilon_b = 52$ (infinite system) is not observed for thin layers. We can conclude that the boundary effects may change very strongly the dispersion characteristics $\varepsilon_{ef}(f)$ near the resonance frequencies.

In the case of thin composite layers $h < h_c$, the renormalised parameter $\varepsilon_b$ used to construct EMT is nearly uniform having almost no dependence on the fibre position $h_0$. Contrary, for thicker materials there are transition layers where $\varepsilon_b$ changes significantly, as shown in Fig. 9. Then, the effective dielectric constant is not uniform either and the existence of the transition layers in thick composites seems as if it may effect the wave propagation changing such measured parameters as transmission, reflection and absorption coefficients. Then, there is a question if this can be a cause in shifting the measured resonance frequency in thick layers, as discussed in Fig. 3. To answer this question, the response from a thick layer ($h >> h_c$) is calculated, dividing it into a number of layers where $\varepsilon_b$ is considered to be uniform. The calculation is done by the matrix method (Abele's method [15,29]). Figure 10 compares the dispersion of the absorption coefficient for a single layer with the matrix constant $\varepsilon = 52$ and for a layered system with $\varepsilon_b$ distributed as shown in Fig. 9. The internal part of the sample with $\varepsilon_b \approx \varepsilon$ is considerably larger than the surface transition layers when $h_c << h$. Then, the variation in $\varepsilon_b$ in these layers is not sufficient to change the total response from the system. It means, that in thick materials with $h >> h_c$ the EMT approach has no need to be modified.

## V. Conclusion

The effective medium theory (EMT) applied to thin composite layers with conducting sticks is developed taking into account the surface displacement currents. The boundary effects are treated within a single-particle approximation, within which it is possible to transform the current distribution at the inclusion to the form similar to that for an infinite system. In this approach, the boundaries can be eliminated considering that the inclusion is embedded in a new matrix with a renormalised dielectric constant. The cross section of this area is of the order of thickness of the transition layer. After this step, the standard procedure to obtain EMT can be used. The dispersion of the effective dielectric constant in thin materials is a function of thickness being significantly different from that for “bulk” materials: the resonance frequency is shifted to higher frequencies and the interval between two resonances is increased.

## Appendix A

In this part, the differential equations completed with the boundary conditions for the vector potential **A** in the system shown in Fig. 1 are obtained. If a fibre is placed in an infinite medium, the field scattered by it is described by the vector potential having only $\mathrm{A}_x$ component. In the case when the surface effects are essential, using one component of the vector potential is in conflict with the condition of continuity of the fields $\mathbf{E}$ and $\mathbf{H}$ at the interface vacuum-dielectric. As it follows from Eq. (2), two components $\mathrm{A}_x$ and $\mathrm{A}_z$ are sufficient to satisfy the continuity condition.[30] The equations for the components $\mathrm{A}_x$ and $\mathrm{A}_z$ are written as

$$\begin{aligned} &\Delta \mathrm{A}_{x1,2,3} + k_{1,2,3}^2 \mathrm{A}_{x1,2,3} = -\mathrm{j}_{1,2,3} \\ &\Delta \mathrm{A}_{z1,2,3} + k_{1,2,3}^2 \mathrm{A}_{z1,2,3} = 0 \end{aligned} . \qquad \text{(A1)}$$

The following boundary conditions are imposed

$$\begin{aligned} &\mu \mathrm{A}_{x2}\big|_{z=0} = \mathrm{A}_{x1}\big|_{z=0}, \qquad \mu \mathrm{A}_{x2}\big|_{z=-h} = \mathrm{A}_{x3}\big|_{z=-h} \\ &\left.\frac{\partial \mathrm{A}_{x2}}{\partial z}\right|_{z=0} = \left.\frac{\partial \mathrm{A}_{x1}}{\partial z}\right|_{z=0}, \quad \left.\frac{\partial \mathrm{A}_{x2}}{\partial z}\right|_{z=-h} = \left.\frac{\partial \mathrm{A}_{x3}}{\partial z}\right|_{z=-h} \end{aligned} \qquad \text{(A2)}$$

$$A_{z2}\big|_{z=0} = A_{z1}\big|_{z=0}, \quad A_{z2}\big|_{z=-h} = A_{z3}\big|_{z=-h}$$

$$\frac{1}{\varepsilon}\left(\frac{\partial A_{x2}}{\partial x} + \frac{\partial A_{z2}}{\partial z}\right)\Bigg|_{z=0} = \left(\frac{\partial A_{x1}}{\partial x} + \frac{\partial A_{z1}}{\partial z}\right)\Bigg|_{z=0}$$

$$\frac{1}{\varepsilon}\left(\frac{\partial A_{x2}}{\partial x} + \frac{\partial A_{z2}}{\partial z}\right)\Bigg|_{z=-h} = \left(\frac{\partial A_{x3}}{\partial x} + \frac{\partial A_{z3}}{\partial z}\right)\Bigg|_{z=-h}$$

where indexes 1, 2, 3 designates areas $z \in [0, \infty+]$, $z \in [-h, 0]$, and $z \in [-\infty, -h]$ respectively, $j_{1,3} \equiv 0$, $j_2 = -j(x)\delta(y)\delta(z+h_0)$, $k_1 = \omega/c$, $k_2 = (\omega/c)\sqrt{\varepsilon\mu}$.

The solution of Eq. (A1) can be presented in the form of convolutions of Green's functions and current density $j(x)$, as it is shown in Eq. (3). The coupled equations (4),(5) are solved using the double Fourier's transformation with respect to variables $x$ and $y$, which yields the differential equations with respect to $z$ for the obtained Fourier's transformations. The integrals in Eqs. (6) and (7) are inverse Fourier's transformations written in the cylindrical co-ordinates with $\rho = \sqrt{x^2 + y^2}$ (known as Sommerfield's integrals [30]).

## Appendix B

The theory developed here requires two types of integrals to be calculated. They are (see Eqs. (6) and (7))

$$\int_0^{+\infty} \frac{f(k, h, h_0, z)}{\Delta_2(k, h)} J_0(k\rho)dk \quad \text{or} \quad \int_0^{+\infty} \frac{g(k, h, h_0, z)}{\Delta_1(k, h)\Delta_2(k, h)} J_0(k\rho)dk \qquad \text{(B1)}$$

where f and g are analytical functions with respect to variable $k$, $\rho = \sqrt{x^2 + y^2}$, functions $\Delta_1$, $\Delta_2$ are determined in (6) and (7). The difficulty in calculating (B1) occurs since in the absence of the dissipation in the medium ($\mathrm{Im}\,\varepsilon = \mathrm{Im}\,\mu \equiv 0$) the functions $\Delta_1(k,h)$ and $\Delta_2(k,h)$ have real zeros, which are located symmetrically in the regions $-k_2 \le k \le -k_1$ and $k_1 \le k \le k_2$.

The equations $\Delta_1(k,h) = 0$ and $\Delta_2(k,h) = 0$ are called dispersion equations of the layered media, which determine the propagation numbers of those waves penetrating inside the medium at sufficient distance. The roots obey the following properties (which are given here without proof): (i) they are positioned symmetrically about $k=0$ (then, only positive $k$ will be considered); (ii) in the absence of dissipation there are $N=(2+[\sqrt{k_2^2-k_1^2}h/\pi])$ different roots in the region $k_1 \le k \le k_2$, where the square brackets designate the integer part; (iii) for $h \to \infty$ this number increases to infinite and the roots tend to occupy the region $k_1 \le k \le k_2$ continuously; (iv) for $h < \pi \big/ \sqrt{k_2^2-k_1^2}$ there is the only root which at $h \to 0$ approaches $k_1$; (v) for $h = \pi n \big/ \sqrt{k_2^2-k_1^2}$, where $n \ge 1$ is any integer number, one of the roots equals $k_1$.

To find real positive roots in the region $k_1 \le k \le k_2$, it is convenient to represent the dispersion equations in the form

$$\Delta_1(k,h) = \left(-\tilde{\gamma}_2^2 + \gamma_1^2\varepsilon^2\right)\sin(\tilde{\gamma}_2 h) + 2\gamma_1\tilde{\gamma}_2\varepsilon\cos(\tilde{\gamma}_2 h) = 0$$

which is equivalent to

$$\left(\gamma_1^2\varepsilon^2 + \tilde{\gamma}_2^2\right)\sin\left(\tilde{\gamma}_2 h + \mathrm{arctg}\left(2\gamma_1\tilde{\gamma}_2\varepsilon \Big/ \left(\gamma_1^2\varepsilon^2 - \tilde{\gamma}_2^2\right)\right)\right) = 0 \qquad \text{(B2)}$$

where $\tilde{\gamma}_2 = \sqrt{k_2^2 - k^2}$, $\gamma_1(k)$ and $\tilde{\gamma}_2(k)$ are real functions of variable $k$ in the considered region. A similar form can be written for $\Delta_2 = 0$. From Eq. (B2) the condition of root existence is obtained

$$\tilde{\gamma}_2 h + \mathrm{arctg}\left(2\gamma_1\tilde{\gamma}_2\varepsilon \Big/ \left(\gamma_1^2\varepsilon^2 - \tilde{\gamma}_2^2\right)\right) = \pi m \qquad \text{(B3)}$$

where $m$ is integer number including zero. For any $m$ there is only one root and the maximum value of $m$ equals to $N=(2+[\sqrt{k_2^2-k_1^2}h/\pi])$ as mentioned above. Eq. (B3) can be solved by a graphical method finding the intersection of lines: $f_1(k) = -\tilde{\gamma}_2 h + \pi m$ and $f_2(k) = \mathrm{arctg}\left(2\gamma_1\tilde{\gamma}_2\varepsilon \Big/ \left(\gamma_1^2\varepsilon^2 - \tilde{\gamma}_2^2\right)\right)$, which is easily realised numerically.

In a standard method, [30] the integrals containing functions with real poles are calculated by integrating along the vertical cuts made from the poles. If the number of poles is large this method becomes unpractical and is of no use. The integration of (B1) can be made by introducing a small dissipation in the system: a small imaginary part appears in $\varepsilon$ (and $\mu$). For example, if the medium has a small conductivity $\sigma$, then $\operatorname{Im}\varepsilon = -i\omega\sigma/4\pi$ (the sign "–" corresponds to time dependence of $\exp(+i\omega t)$). As a result the poles shift in the complex plane below the real axis and the integrals can be calculated in a usual way (as classical). The challenge now is to calculate the limit of the integrals when the dissipation approaches zero. For this purpose, the integrals of Cauchy type are considered [31]

$$\mathrm{F}(q) = \frac{1}{2\pi i}\int_C \frac{\Psi(k)dk}{k-q} \qquad \text{(B4)}$$

where $\Psi(k)$ is a continuous complex function, $C$ is an arbitrary path in the complex plain $\mathrm{P}$, and $q \in \mathrm{P}$ is a complex number. The limit of $\mathrm{F}(q)$ when the point $q$ approaches a point $k^*$ on the path $C$ can be found by means the Plemelj formulas (or, less often, the Sokhotski formulas)

$$\mathrm{F}^+(k^*) = \mathrm{F}(k^*) + \frac{1}{2}\Psi(k^*),$$

$$\mathrm{F}^-(k^*) = \mathrm{F}(k^*) - \frac{1}{2}\Psi(k^*). \qquad \text{(B5)}$$

Here $\mathrm{F}^+(k^*)$ and $\mathrm{F}^-(k^*)$ are the "left" and "right" limits when the point $q$ approaches the point $k^*$ from the "left" and "right" with respect to the integral path. The function $\mathrm{F}(k^*)$ is defined by the following equation

$$\mathrm{F}(k^*) = \frac{1}{2\pi i}\int_C \frac{\Psi(k)-\Psi(k^*)}{k-k^*}dk + \frac{1}{2}\Psi(k^*) + \frac{\Psi(k^*)}{2\pi i}\mathrm{Ln}\frac{b-k^*}{a-k^*} \qquad \text{(B6)}$$

where $\mathrm{Ln}((b-k^*)/(a-k^*)) = \ln\left|((b-k^*)/(a-k^*))\right| + i\arg((b-k^*)/(a-k^*))$ is the principal value of a logarithm, and $a, b$ are the ends of the path. In the absence of the dissipation when $k^*$ is a real root of $\Delta_1(k,h)=0$ or $\Delta_2(k,h)=0$, the integrals in Eq. (B1) can be rewritten in the form of the Cauchy integrals

$$\int_{a_i}^{b_i} \frac{\mathrm{f}(k,h,h_0,z)}{\Delta_2(k,h)} \mathrm{J}_0(k\rho)dk = \int_{a_i}^{b_i} \frac{\mathrm{f}(k,h,h_0,z)(k-k_i^*)/\Delta_2(k,h)}{k-k_i^*} \mathrm{J}_0(k\rho)dk \,. \tag{B7}$$

Here $\Psi(k) = \mathrm{f}(k,h,h_0,z)(k-k_i^*)/\Delta_2(k,h)$ is a continuous function of the variable $k$, the parameter $k_i^*$ is the $i$'th real root of the dispersion equation, and $(a_i, b_i)$ is the subinterval containing $k_i^*$. The value of $\Psi(k^*)$ is found via residue: $\Psi(k_i^*) = \mathrm{res}[\mathrm{f}(k,h,h_0,z)/\Delta_2(k,h)]_{k=k_i^*}$. The integrals in Eq. (B1) can be calculated by dividing the integration path into intervals: $(0,k_1),...,(a_i,b_i),...,(k_2,+\infty)$ where the integration in the intervals $(a_i,b_i)$ is carried out as explained above. In Eq. (B5) the «right» limit $\mathrm{F}^-(k_i^*)$ has to be used since in the case of dissipation the roots shift below the real axis.

**Appendix C**

In this Appendix, the approximation used to obtain equations (14) and (16) is discussed. Let us consider equation (13) for the current density, which is written as

$$\frac{\partial^2}{\partial x^2}\left( \mathrm{j}(x) + \frac{(\mathrm{G}_1 * \mathrm{j})}{Q_1} \right) + k_b^2 \left( \mathrm{j}(x) + \frac{(\mathrm{G}_2 * \mathrm{j})}{Q_2} \right) = -\frac{i\omega e}{4\pi Q_1} E_{0x} \tag{C1}$$

where the following notations are used: $\mathrm{G}_1 = i\,\mathrm{Im}\left(\mathrm{G}_{x2} + \tilde{\mathrm{U}}_2\right)$, $\mathrm{G}_2 = i\,\mathrm{Im}\left(\mathrm{G}_{x2}\right)$, and the wave number $k_b^2 = k_2^2 Q_2 / Q_1$. As it follows from Eq. (11) the functions $\mathrm{G}_1$ and $\mathrm{G}_2$ are proportional to

$$\mathrm{G}_{1,2} \sim i \frac{\sin(k_b x)}{\sqrt{x^2 + r_0^2}} \tag{C2}$$

The order of $Q_1$ and $Q_2$ is estimated from Eqs. (11),(12)

$$Q_{1,2} \propto \int_{-l/2}^{+l/2} \frac{dx}{\sqrt{x^2 + r_0^2}} \approx 2\ln(l/r_0) \tag{C3}$$

Equation (C1) can be represented as a differential equation with respect to the function $\mathrm{j}(x) + (\mathrm{G}_1 * \mathrm{j})/Q_1$

$$\frac{\partial^2}{\partial x^2}\left(\varphi(x)+\frac{(\Gamma_1 * \varphi)}{Q_1}\right)+k_b^2\left(\varphi(x)+\frac{(\Gamma_1 * \varphi)}{Q_1}\right)=-\frac{i\omega e}{4\pi Q_1}E_{0x}+k_b^2\left(\frac{(\Gamma_1 * \varphi)}{Q_1}-\frac{(\Gamma_2 * \varphi)}{Q_2}\right) \qquad \text{(C4)}$$

The general solution of Eq. (C4) can be written in the form of an integral equation

$$\varphi(x)=-\frac{i\omega e}{4\pi Q_1 k_b^2}E_{0x}+A\sin(k_b x)+B\cos(k_b x)-\frac{(\Gamma_1 * \varphi)}{Q_1}+k_b\int_{-l/2}^{x}\sin(k_b(x-s))\Phi(s)ds \qquad \text{(C5)}$$

where $\Phi(s)=((\Gamma_1 * \varphi)/Q_1-(\Gamma_2 * \varphi)/Q_2)$, $A$ and $B$ are constants which are defined from the boundary conditions $\varphi(-l/2)=\varphi(l/2)=0$. The first three terms can be used to construct the zero-order solution

$$\varphi_0(x)=-\frac{i\omega e}{4\pi Q_1 k_b^2}E_{0x}+A\sin(k_b x)+B\cos(k_b x). \qquad \text{(C6)}$$

The next-order terms can be found from a standard iteration procedure

$$\varphi_n(x)=\varphi_0(x)+\int_{-l/2}^{l/2}\tilde{\Phi}(x,\theta)\,\varphi_{n-1}(\theta)d\theta;\ n=1,N \qquad \text{(C7)}$$

where $\tilde{\Phi}(x,\theta)$ is the kernel of the total linear integral operator

$$\tilde{\Phi}(x,\theta)=-\frac{1}{Q_1}\Gamma_1(x-\theta)+k_b\int_{-l/2}^{x}\sin(k_b(x-s))\left(\frac{\Gamma_1(s-\theta)}{Q_1}-\frac{\Gamma_2(s-\theta)}{Q_2}\right)ds \qquad \text{(C8)}$$

The constants $A$ and $B$ are found from the boundary conditions $\varphi_n(-l/2)=\varphi_n(l/2)=0$ at the final stage of the iteration procedure for a fixed $n=N\geq 1$. Since the integral operator is linear the equation for $A$ and $B$ form a linear system. The solution for $\varphi_0$ ($N=0$) has the form

$$\varphi_0(x)=\frac{i\omega e E_{0x}}{4\pi Q_1 k_b^2}\frac{(\cos(k_b x)-\cos(k_b l/2))}{\cos(k_b l/2)} \qquad \text{(C9)}$$

Equation (C9) gives singularity at the resonance frequency when $\cos(k_b l/2)=0$, which can be eliminated using the next iterations. Since the kernel $\tilde{\Phi}(x,\theta)$ involves parameters $1/Q_1, 1/Q_2 \propto 1/2\ln(l/r_0)$, the last term in the N-th iteration series is of the order of $(1/2\ln(l/r_0))^{N+1}$. In the present case, the iteration parameter is sufficiently small: taking $l/2 = 0.4$ cm and $r_0 = 4 \cdot 10^{-4}$ cm gives $1/2\ln(l/r_0) \approx 0.07$. Then, the higher-order iterations will introduce small changes in the effective parameter $k_b$ and the resonance frequency (but can change the current at the resonance considerably, which is not important in this case) and thus the form factors $Q_1$ and $Q_2$ properly define the modified effective parameters. The zero-order iteration corresponds to neglecting the imaginary parts in equation (13).

## Figures

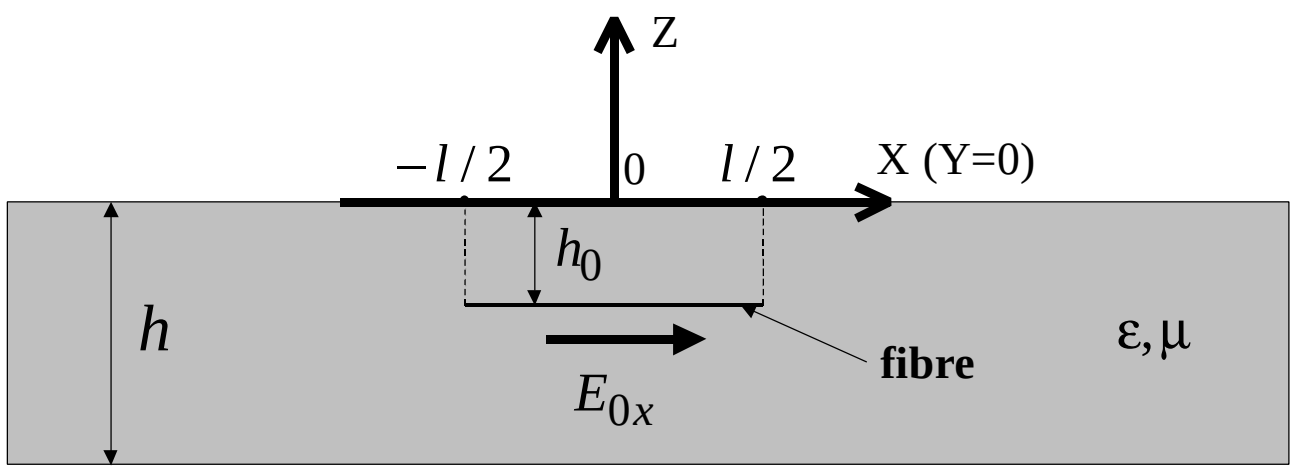

Fig. 1. Geometry of a single-particle problem: a dialectic layer with a conducting fibre.

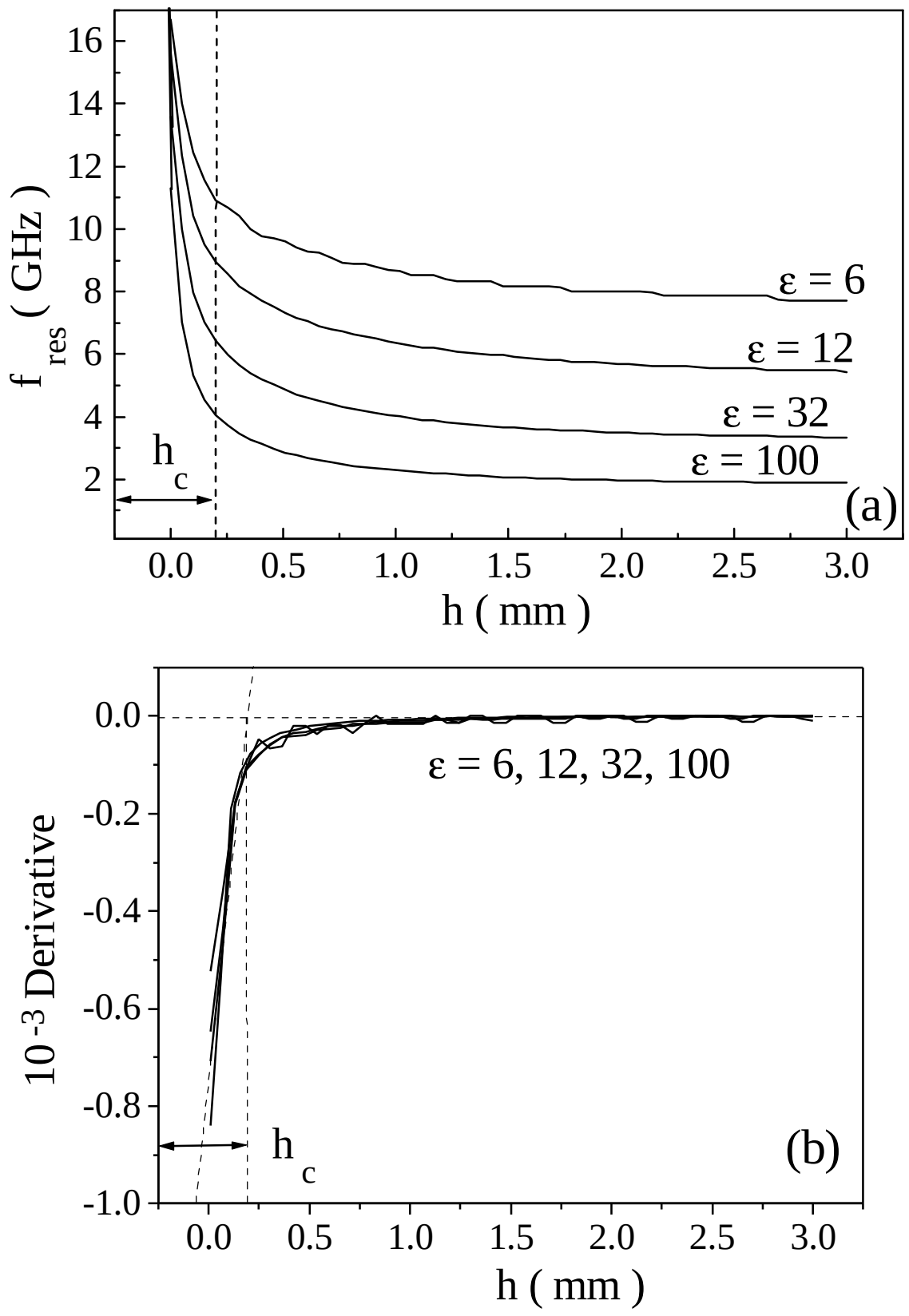

Fig. 2. Resonance characteristics of a dielectric layer with a conducting fibre. In a), resonance frequency $f_{res}$ as a function of the layer thickness *h* for different dielectric matrixes, in b) derivative $\partial f_{res} / \partial h$ versus *h*, which defines the characteristic transition layer thickness $h_c$.

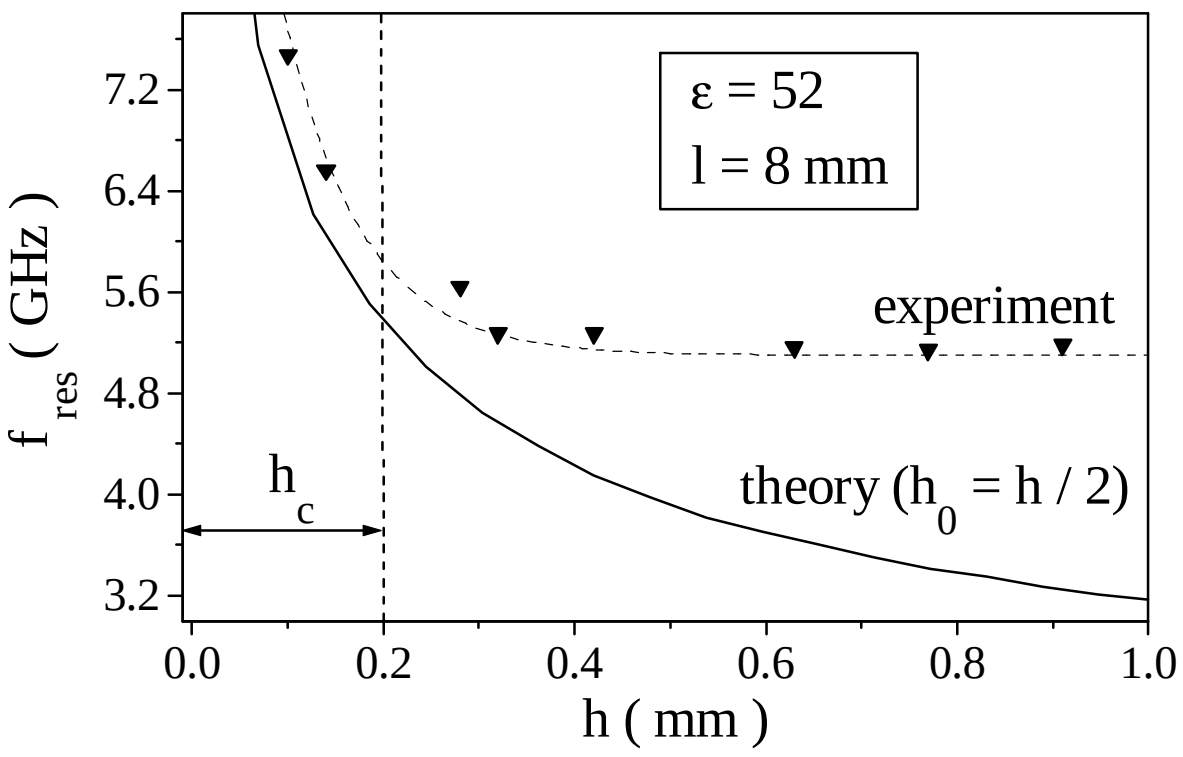

Fig. 3. Plots of $f_{res}$ versus $h$, comparison of theory (solid curve) and experiment (dashed curve).

$\varepsilon = 52$, $l = 8$ mm, $h_0 = h/2$

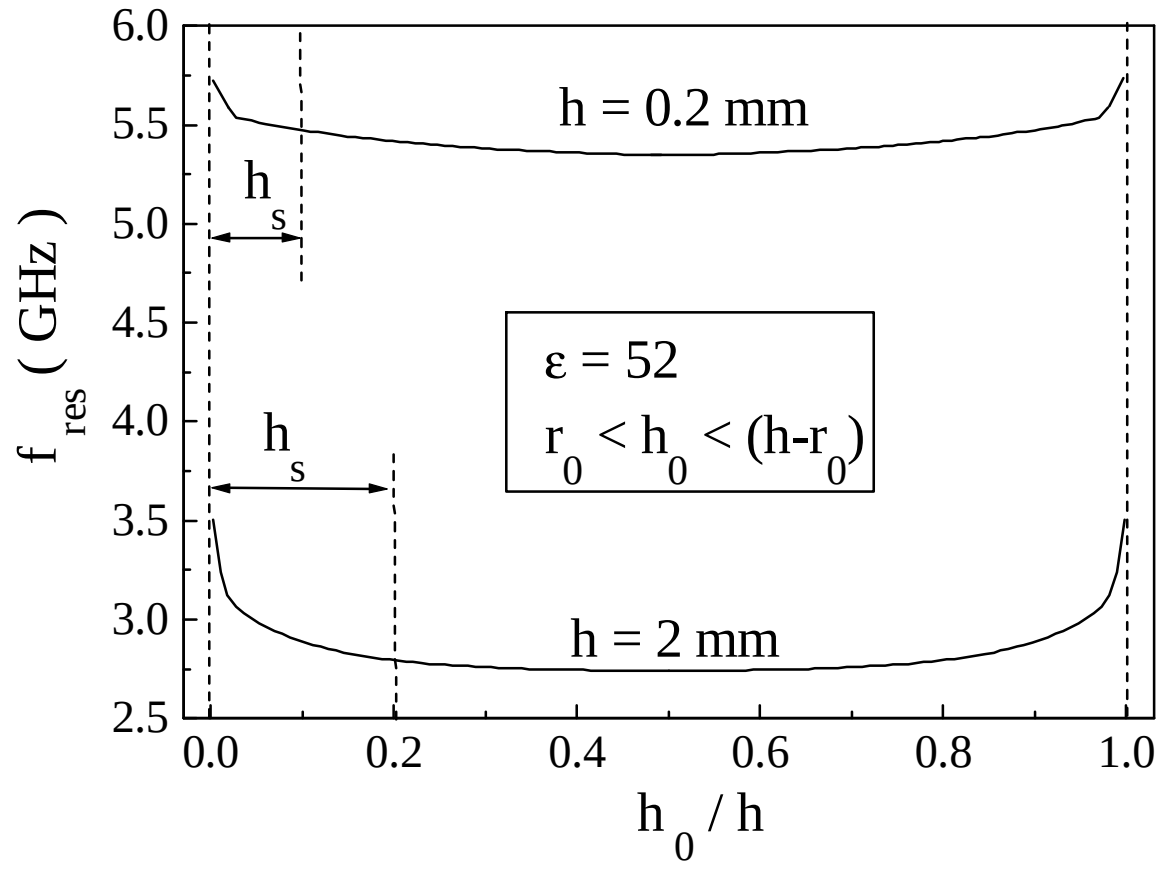

Fig. 4. Resonance frequency $f_{res}$ as a function of the fibre position $h_0 / h$ for two layer thickness

$h = 0.2$ mm ($\sim h_c$) and $h = 2$ mm ($>> h_c$).

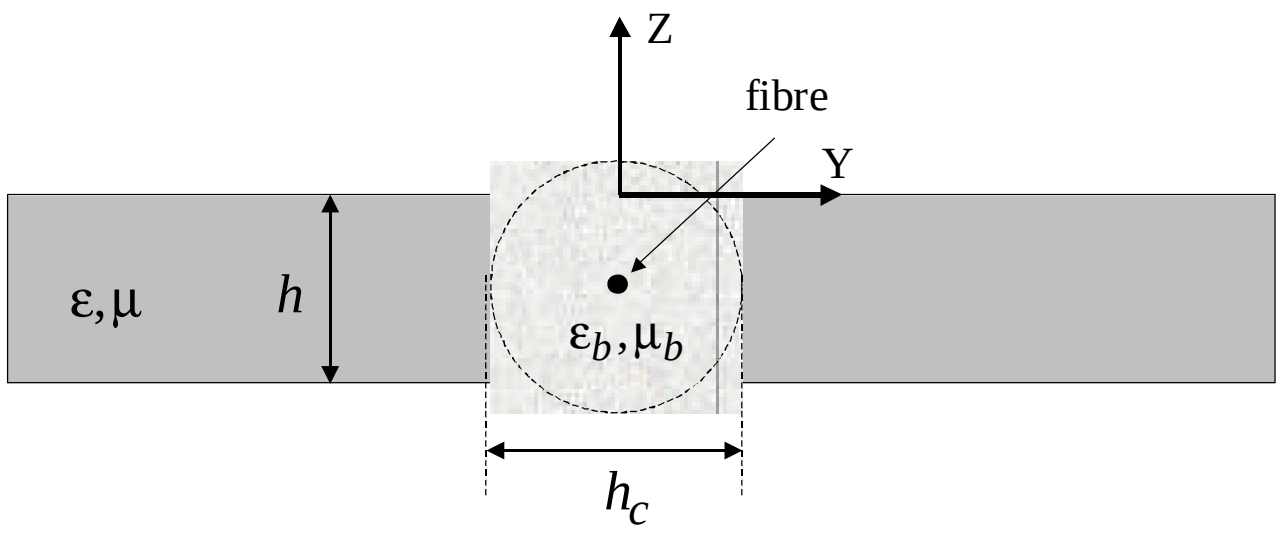

Fig. 5. Structure of a thin layer when the effect of boundaries is replaced by a new medium around the fibre.

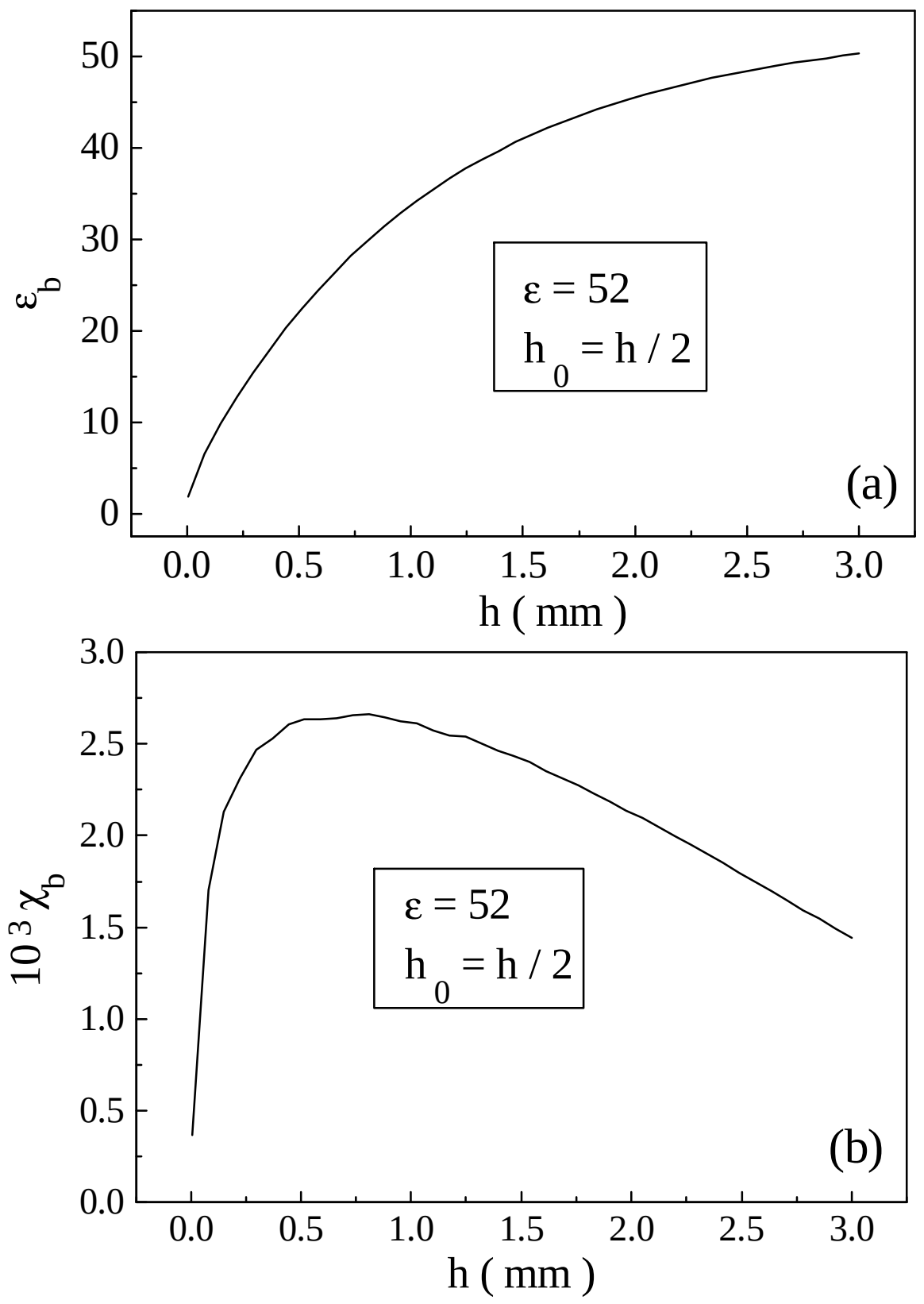

Fig. 6. Renormalised matrix parameters $\varepsilon_b$ (in (a)) and $\chi_b = (\mu_b - 1)/4\pi$ (in (b)) as function of *h*.

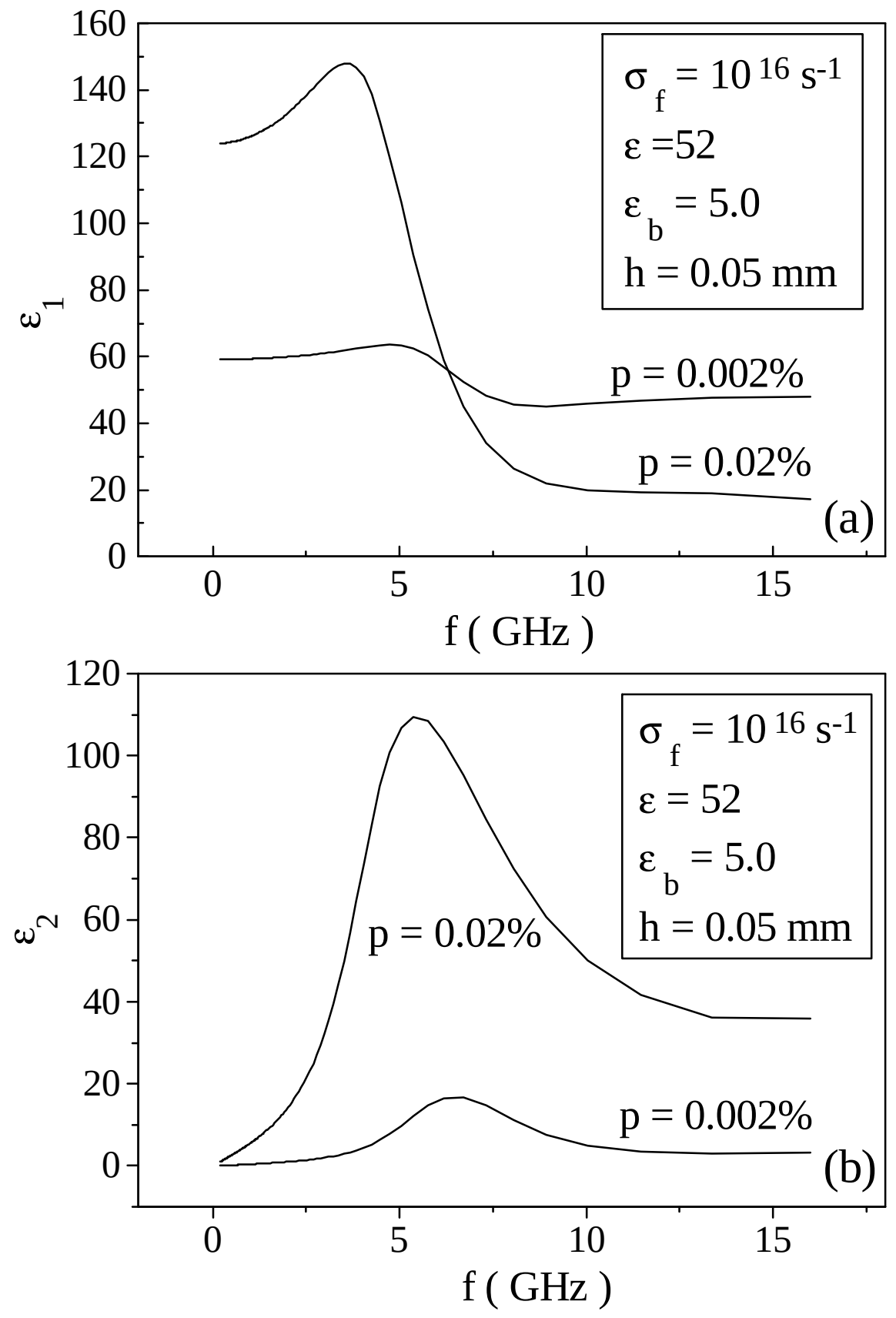

Fig. 7. Effective dielectric constant $\varepsilon_{ef} = \varepsilon_1 + i\varepsilon_2$ as a function of frequency for two concentrations 0.002% and 0.02%, $\varepsilon = 52$, $\varepsilon_b = 5$, $h = 0.05$ mm. Real part $\varepsilon_1$ is in (a) and imaginary part $\varepsilon_2$ is in (b). The aluminium coating had the semicontinuos structure since the effective fibre conductivity $\sigma_f$ was less than in the case of the all-metal inclusions and it is taken to be equal to $10^{16} s^{-1}$.

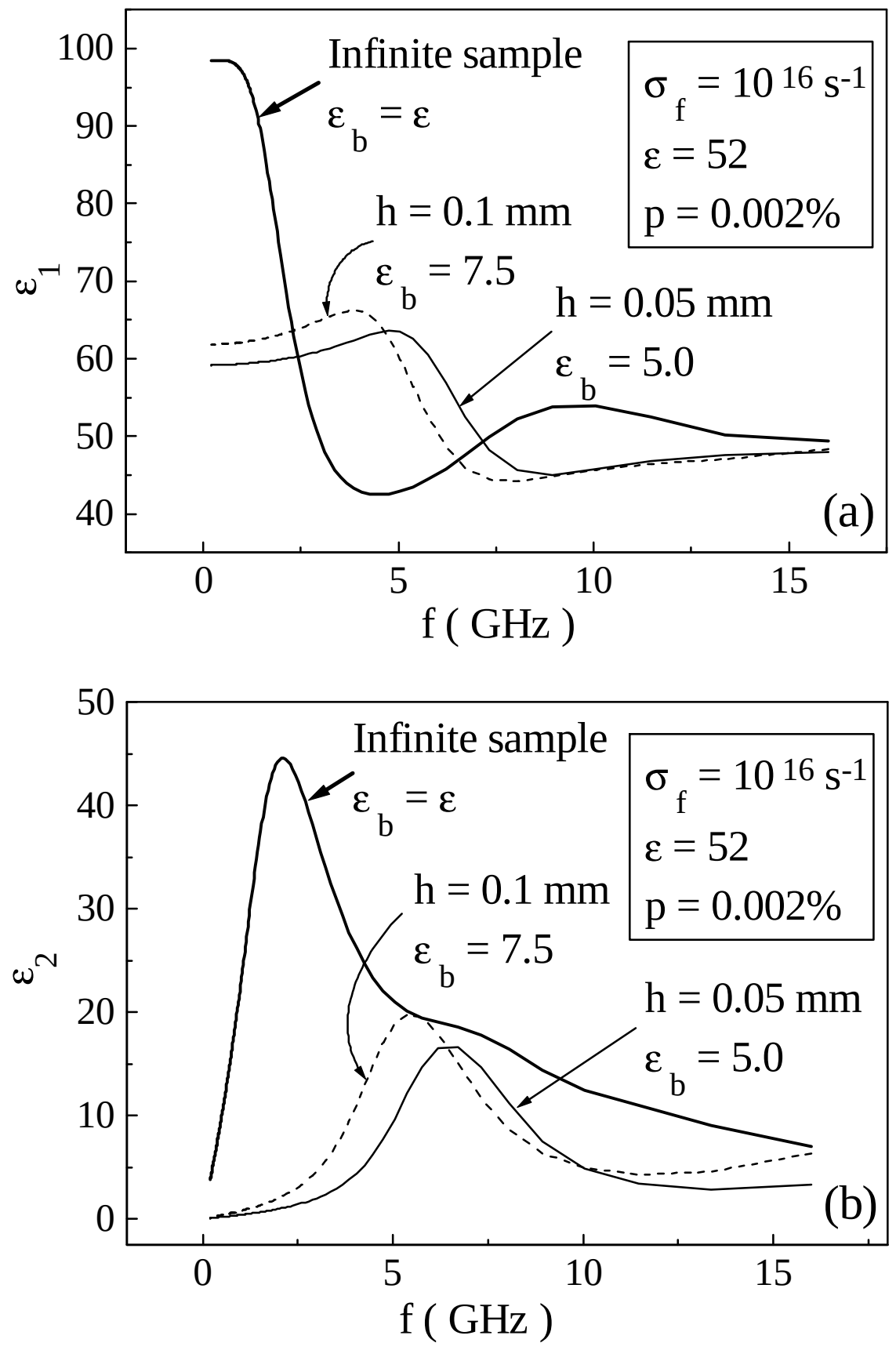

Fig. 8. Effective dielectric constant $\varepsilon_{ef} = \varepsilon_1 + i\varepsilon_2$ as a function of frequency for different thickness *h*. $\varepsilon_1$ is in (a) and $\varepsilon_2$ is in (b), $p = 0.002\%$.

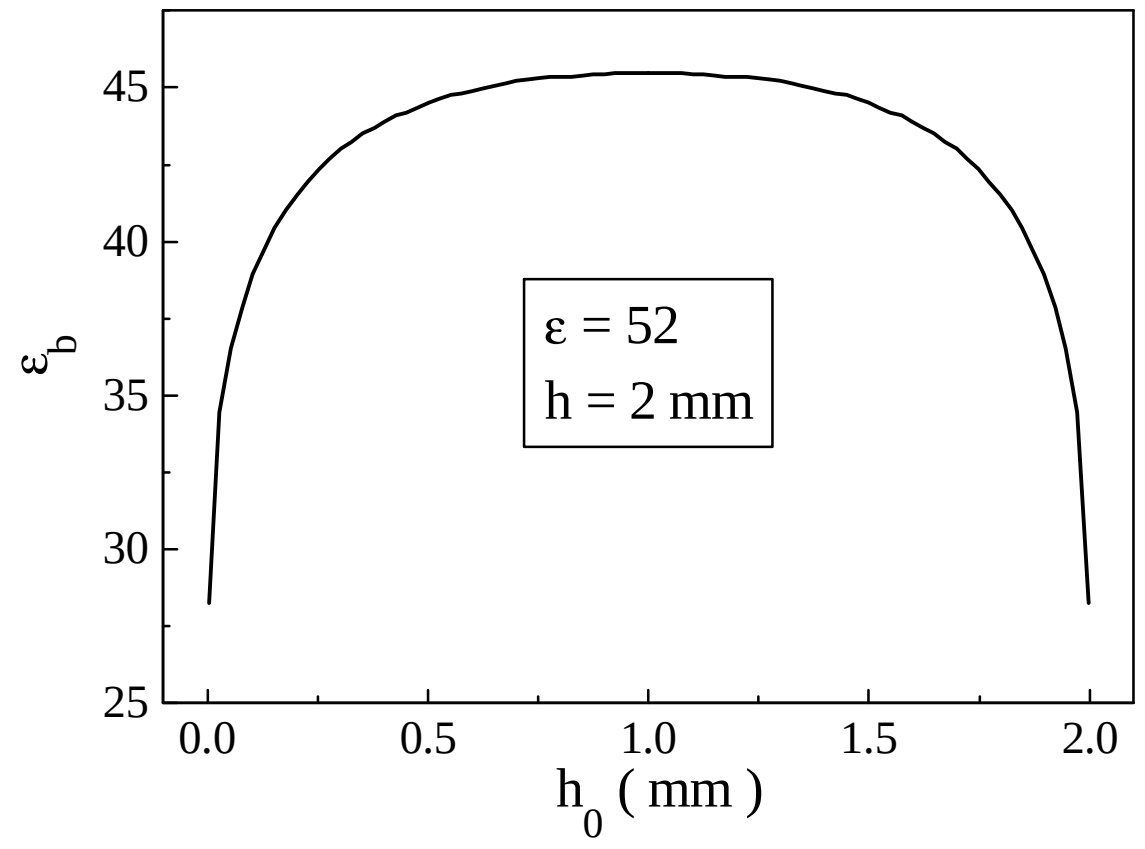

Fig. 9. Renormalised matrix parameters $\varepsilon_b$ as a function of the fibre position $h_0 / h$ for $h = 2$ mm (>> $h_c$).

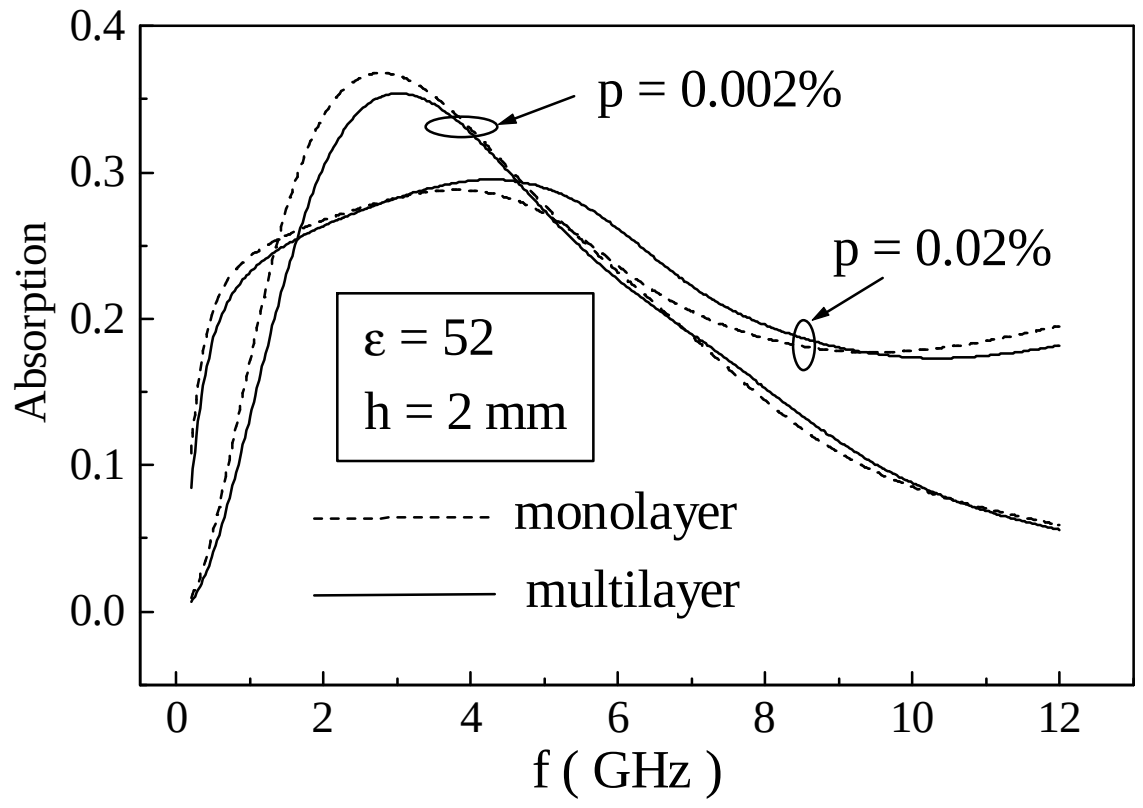

Fig. 10. Effective dielectric response (coefficient of absorption) from a thick composite layer ($h = 2$ mm (>> $h_c$)). Solid curve: multilayer system. Dashed curve: uniform single layer.